\documentstyle[aps,prl,multicol,epsfig,array]{revtex}

\def\be{\begin{equation}}
\def\ee{\end{equation}}

\begin{document}


\title{An extremal model for amorphous media plasticity}
\author{Jean-Christophe Baret, Damien Vandembroucq and St\'ephane Roux}
\address{Unit\'e Mixte CNRS/Saint-Gobain ``Surface du Verre et Interfaces''\\
39 Quai Lucien Lefranc, 93303 Aubervilliers cedex, FRANCE}

\maketitle 
\begin{abstract} 
An extremal model for the plasticity of amorphous 
materials is studied in a simple two-dimensional anti-plane geometry.  
The steady-state is analyzed through numerical simulations.  Long-range
spatial and temporal correlation in local slip events are shown to
develop leading to non-trivial and highly anisotropic scaling laws.  
In particular, the plastic
strain is shown to statistically concentrate over a region which tends
to align perpendicular to the displacement gradient.  
By construction, the model can be seen as giving rise to a depinning 
transition, the threshold of which (i.e. the macroscopic yield stress)
also reveal scaling properties reflecting the localization of the
activity.  
\end{abstract}
\pacs{PACS numbers }

\begin{multicols}{2}

In contrast with crystalline solids, amorphous materials display a
plasticity which cannot be resumed to the motion of well identified
defects such as dislocations.  Consequently, the microscopic
description of amorphous plasticity is still lacking a consistent
framework.  Recent studies
\cite{BulatovArgon94a,BulatovArgon94b,BulatovArgon94c,FalkLangerPRE98}
have focussed on the fact that global plastic deformation is mostly
due to local rearrangements.
Starting from a molecular dynamics study of a bidimensional
Lennard-Jones glass and measurements of the mechanical response under
shear stress, Falk and Langer\cite{FalkLangerPRE98} introduced the
notion ``Shear Transformation Zones'' (STZ) having a bistable
character to build a mean field theory of plastic deformation in an
amorphous material.  Initially drawn by Bulatov and Argon
\cite{BulatovArgon94a,BulatovArgon94b,BulatovArgon94c} for amorphous
solids materials this approach can be extended to granular materials
or dense suspensions\cite{Lemaitre01a,Lemaitre01b}.  In the following
we study a minimal model of plastic deformations in disordered
media. This model has been proposed in the early eighties to describe
the fault self-organization in seismic
regions\cite{ChenBakPRA91,SornettePRL93} where mostly the
``avalanche'' properties of this model were studied in order to
compare with the observed power-law distribution of seismic events
(Gutenberg-Richter law). In \cite{SornettePRL93}, a quenched random
distribution of local threshold stress was introduced which allowed
for a mapping onto a random polymer problem in the limit of vanishing
stress drop.  The analysis we propose focuses on the scaling feature
of the spatio-temporal organization of the local slip events and on
the stress-strain characterization.

We consider a bidimensional material submitted to anti plane shear
stress. The elastic component of the displacement $u_z(x,y)$ is thus
solution of a Laplace equation $\nabla^2 u_z =0$. The material is
discretized on a regular lattice, the axis of which are oriented at 45
degrees from the displacement gradient direction.  The elastic modulus
is assumed to be uniform but we impose a spatially frozen disorder for
the onset of slip at a local level. After a local slip, we renew the
threshold stress where the slip occured from a random
distribution. Bi-periodic boundary conditions are implemented for the
stress and the strain, whereas a discontinuity is imposed on the
displacement along the $y$ axis.  We choose an extremal dynamics: the
external load $\Sigma$ is adjusted at each step so that only one bond
is at the plastication threshold. In the spirit of
Ref.\cite{FalkLangerPRE98}, this corresponds to a structural
rearrangement of a ``shear transformation zone''. The latter induces a
displacement discontinuity along the bond and a random modification of
the plastication threshold. The local stress is redistributed over the
material according to the elastic response function. The local stress
on a bond $i$ is $\sigma_i=\Sigma+\sigma_i^o$ where $\Sigma$ is the
macroscopic stress. After a slip $\Delta u_j$ at bond $j$, $\sigma_i^o$ is
adjusted to $\sigma_i^o=\sigma_i^o+\Delta u_j
G(\underline{x_i}-\underline{x_j})$.
Apart from the periodicity imposed by the boundary conditions, this
function $G(\underline{x})$ is long ranged, decreasing as
$G(\underline{x}) \propto |\underline{x}|^{-2}$. Moreover, due to the
shear boundary condition, the stress redistribution is anisotropic. In
the longitudinal $x$ direction, bonds are loaded while in the
transverse $y$ direction they are unloaded. This anisotropy is one
distinct feature of the model leading to the localization effect to be
described below.

The local plastication thresholds $\gamma(x,y)$ are randomly chosen
according to a uniform distribution between 0 and 1. Since the elastic
modulus is uniform, the stress redistribution is computed once and for
all {\it via} a conjugated gradient algorithm, for a local slip of
unit magnitude.  This effective Green function is then simply
translated to the location of the slip event and its amplitude is
scaled by the slip magnitude (also assumed to be random).

The behavior of this model is of the pinning/depinning type. The
plastication criterion of an individual bond can be written
$\Sigma^{ext} > \gamma(x,y) - \sigma^{el}(x,y)$ where $\sigma^{ext}$
is the external shear stress, $\gamma(x,y)$ the local plastication
threshold and $\sigma^{el}(x,y)$ the local stress component due to
elastic stress redistribution from previous plastication events. Following an
extremal dynamics, we select at time $t$ the current ``weakest site''
$(x^*,y^*)$ such that $\sigma_c(t)= \gamma(x^*,y^*) -
\sigma^{el}(x^*,y^*)=\min_{(x,y)} [ \gamma(x,y) - \sigma^{el}(x,y)
]$. The maximum over time $\sigma^*=\max_t \sigma_c(t)$ corresponds to
the macroscopic yield stress.  From this signal we can reconstruct the
evolution of the system subjected to a constant load: when submitted
to an external shear stress lower than $\sigma^*$ the plate deforms
plastically before  blocking in a jammed state. For values
above the yield stress, the system flows indefinitely.

Such pinning systems have been extensively studied over the recent
years. They have been used to describe front motion in a disordered
environment in the context of wetting\cite{Rolley}, magnetic domain
walls \cite{Lemerle}, fluid invasion in porous media \cite{Wilkinson},
crack propagation \cite{Schmitt1,TGRPRE98,Ramanathan}; or the behavior
of `periodic systems', \emph{e.g.}  vortex lattices \cite{Blatter} or
charge density waves (CDW) \cite{Gruner}.

\begin{figure}
\begin{center}
\epsfig{file=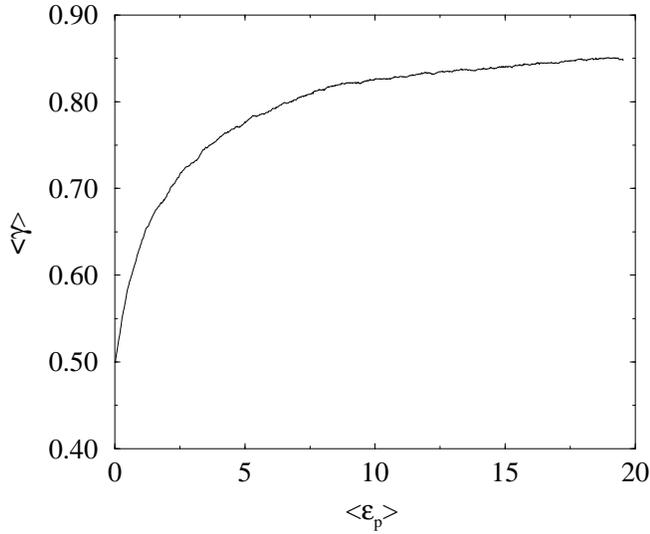, width=0.4\textwidth,angle=-90}
\end{center}
\caption{Evolution of the mean plastication threshold $\langle
\gamma(x,y) \rangle$ during the transient regime of a first loading
process. The increase of $\langle \gamma(x,y) \rangle$ can be
interpreted as a hardening effect.}
\label{hardening}
\end{figure}

Beyond the yielding transition, this simple model exhibits another
characteristic feature of plasticity: hardening ({\it i.e. } increase of the yield stress with the plastic strain). In crystalline
solids, the hardening behavior is due to the entanglement of
dislocation loops. In the present case, after a first loading, we
observe an increase of the elastic limit. However the mechanism for
this hardening effect is here of a pure statistical nature.  During
the loading process, the weakest sites are progressively decimated.
Then the plastic threshold $\gamma(x,y)$ is renewed. The new threshold
is in average larger than the previous one.  This introduces a
systematical bias. When submitted for the first time to a loading
process, the distribution of these local plastication thresholds
evolves to eventually reach a stationary state. On Fig. \ref{hardening},
we show the evolution of the mean plastication threshold $\langle
\gamma(x,y)\rangle$ during loading. 
The asymptotic steady distribution seems not reached yet on the figure.
 This hardening effect thus
corresponds to a progressive reinforcement of the weakest regions.

\begin{figure}
\begin{center}
\epsfig{file=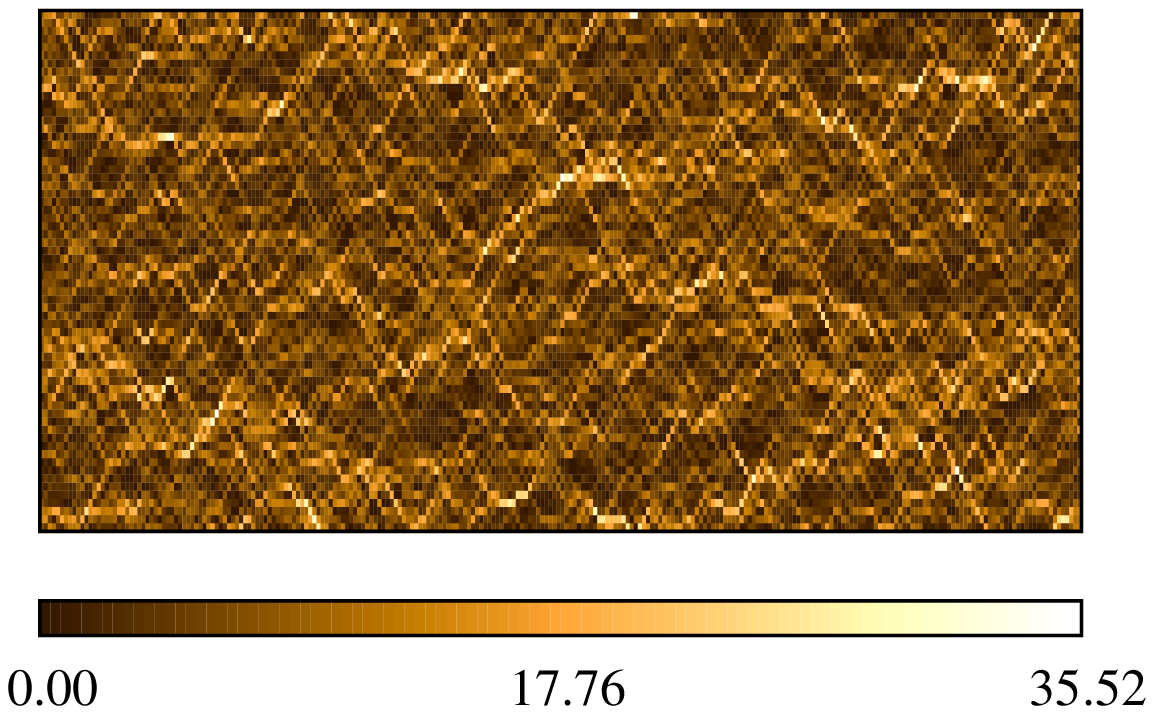, width=0.4\textwidth,angle=0}
\epsfig{file=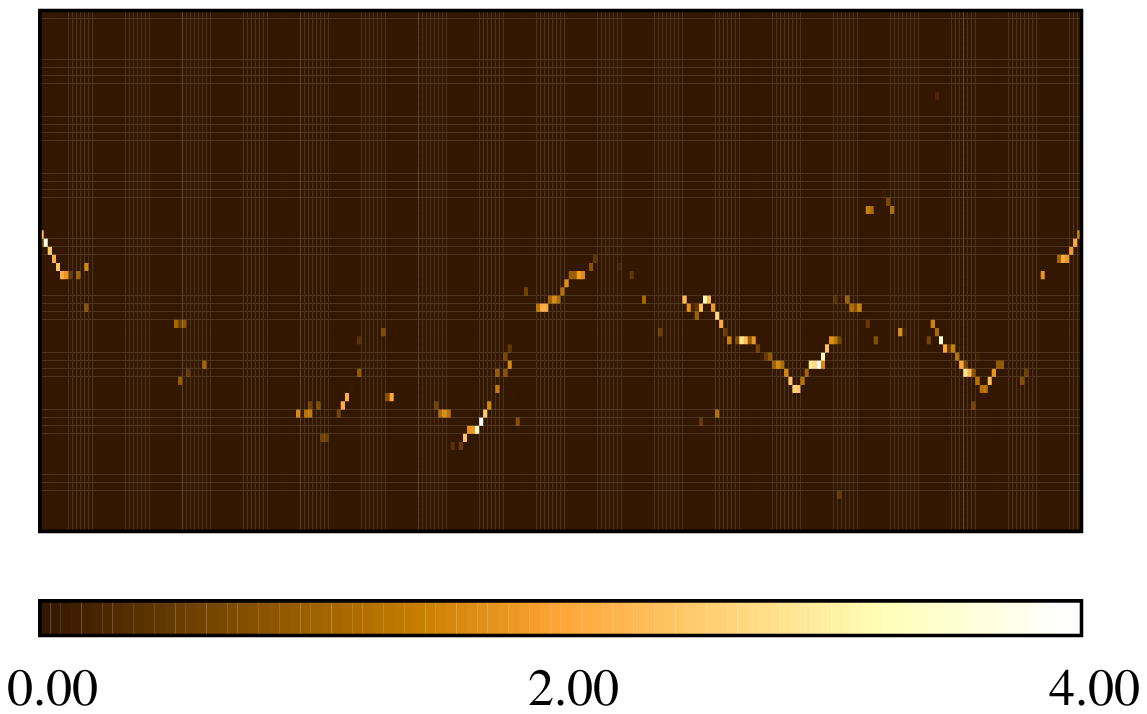, width=0.4\textwidth,angle=0}
\end{center}
\caption{Map of the relative displacement field obtained on a system
$128\times64$ after 800 000 times steps (above). The diffuse
localization corresponds to the succesive development of anisotropic
structures elongated in the longitidinal direction. Focusing on a
finite time window (450 time steps, below) allows to reveal an
individual structure.}
\label{activitymap}
\end{figure}

\begin{figure}
\begin{center}
\epsfig{file=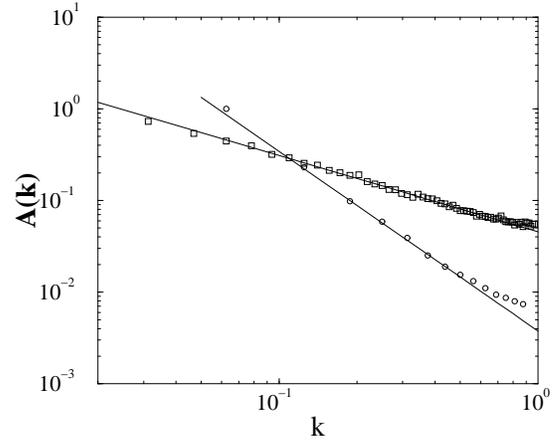, width=0.4\textwidth,angle=0}
\end{center}
\caption{Power spectra of the plastic strain $\varepsilon_p$ for $k_x=0$ (circles) and $k_y=0$ (squares). The lines indicate power law behaviors corresponding to roughness exponents $\zeta_\parallel= -0.09$ and $\zeta_\perp = 0.50$}
\label{figpowerspec}
\end{figure}

On Fig. \ref{activitymap} (above) we show a map of the cumulative plastic strain
for a system of size $128\times64$ after $8 10^5$ time steps.  We see
clearly that the plastic strain is non uniform: it is localized within
regions elongated along the $x$ direction. 
Focusing on the plastic deformation taking place within a finite time
window, we show on the same figure (below) the appearance of an
individual localized structure.
 To characterize quantitatively this spatial distribution, we studied
the pair correlation function of the plastic strain
$\varepsilon_{p}(x,y)$ through Fourier transforms of the strain map
averaged over time.  
We found that
the projection of the plastic strain along the $x$ or $y$
axis, $\varepsilon_\parallel(x)=\langle\varepsilon_p(x,y)\rangle_y$
and $\varepsilon_{\perp}(y)=\langle\varepsilon_p(x,y)\rangle_x$ are
self-affine profiles with roughness exponents $\zeta_\parallel\approx
-0.09$ and $\zeta_\perp\approx 0.50$.  Figure \ref{figpowerspec} shows
the power spectra of $\varepsilon_p$ for $k_x=0$ and $k_y=0$, where
the power-law behaviors give directly the cited roughness exponents.

Such a scaling behavior which characterizes the steady state fluctuations of the 
cumulative strain allows to analyze the time evolution of
the plastic flow.  Let us consider two local slip events 
separated by a time lapse $\tau$, and record their distance along the $x$
and $y$ direction, noted respectively $d_\parallel$ and $d_\perp$. 
Averaging over time (at fixed $\tau$), the probability distribution
function of these distances $p(d,\tau)$ reveal two characteristic ``correlation
lengths'', $\xi_\parallel$ and $\xi_\perp$, below which $p$ is constant,
and above which $p$ decays as a power-law with an exponent
$\alpha_\parallel$ or $\alpha_\perp$ respectively.  
Varying the time lapse $\tau$, we observe that 
\begin{equation}
\xi_\parallel(\tau)\propto \tau^{1/z_\parallel}\qquad
\xi_\perp    (\tau)\propto \tau^{1/z_\perp}
\end{equation}
Exploiting the self-affine nature of the cumulative plastic strain,
and using a result obtained for other extremal models of depinning, we
can relate the two dynamic exponents to the roughness
exponents\footnote{When $\zeta <0$, an effective value of $\zeta^{eff}=0$
should be read in this formula}:
\begin{equation}
z_\parallel=1+\zeta_\parallel \qquad
z_\perp    =1+\zeta_\perp
\end{equation}
The numerical values of the $z$ exponents are consistent with these identities.

The difference in scaling in the $x$ and $y$ direction can be accounted
for through a power-law relating both directions.  Indeed, the
correlation lengths are related through $\xi_\perp\propto 
\xi_\parallel^\beta$ with $\beta=z_\parallel/z_\perp\approx 0.65$. 
Moreover, looking at the mean value of $d_\perp$ for a prescribed value
of $d_\parallel$ also reveal the same power-law $d_\perp\propto
d_\parallel^\beta$, with $\beta\approx 0.65$.

\begin{figure}
\begin{center}
\epsfig{file=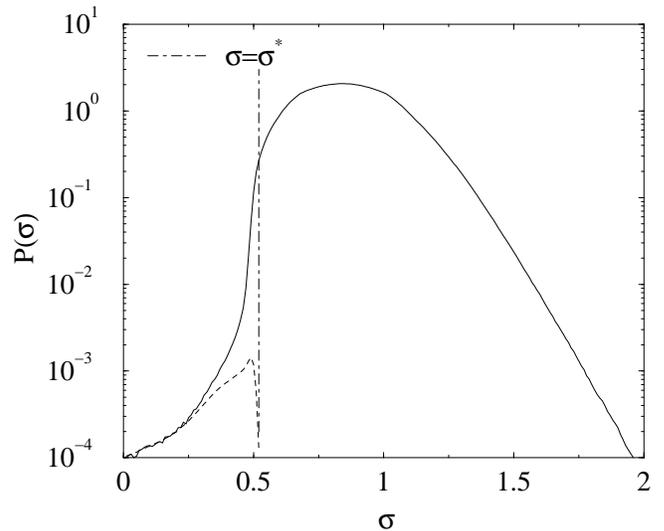, width=0.4\textwidth,angle=-90}
\end{center}
\caption{Distribution of the individual site plastication thresholds
$\sigma_i$ (line) (all sites, all times) and of the current plate plastication thresholds
$\sigma_c$ (dotted line) (active site, all times). The vertical line indicates the position of
the yield stress $\sigma^*$.
}
\label{fiandfc}
\end{figure}

Let us focus now on the depinning stress distribution.  On
Fig. \ref{fiandfc} we show the distribution of the local plastication
stresses, $\gamma(x,y)-\sigma^{el}(x,y)$ (at all sites and all times)
and of the current plastication stresses, $\sigma_c(t)$. The maximum
of the latter over time corresponds to the macroscopic yield stress
$\sigma^*$. We clearly see that the yield stress separates two
distinct regions. Stresses larger than $\sigma^*$ are approximately
distributed according to a normal law. Stresses lower than $\sigma^*$
are however distributed according to a power law of the argument
$(\sigma^*-\sigma)$. As above suggested the fraction of sites such
that $\sigma < \sigma^*$ can be thought of as a population of
potential active sites and hence may be interpreted as potential
STZ. Let us emphasize that these STZ are not postulated but emerge
naturally within the model.

\begin{figure}
\epsfig{file=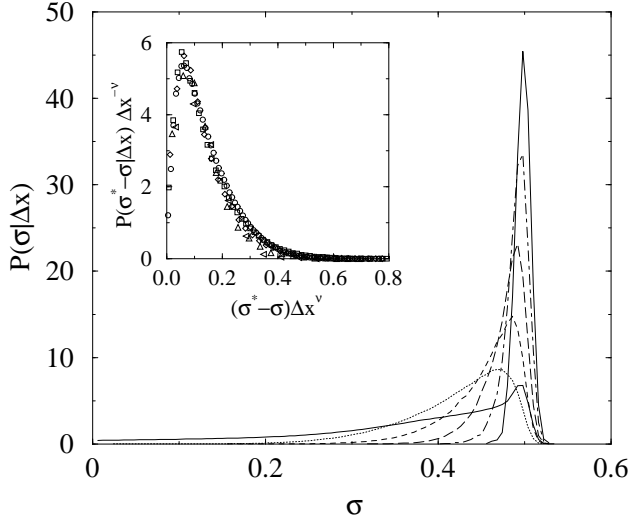,width=0.4\textwidth,angle=-90}
\caption{Distribution of depinning stress (bold) and contributions
conditionned by the distances (2,4,8,16,32) between consecutive active
sites. The tail of the distribution corresponds to very short jumps
and is very sensitive to the details of the random threshold
distribution. The contributions obtained for increasing distances
between consecutive active sites present the same trend: the larger
the jump, the closer the mean force to the threshold and the narrower
the distribution. After rescaling (inset) these distributions collapse
onto a single master curve.}
\label{stress-distribution}
\end{figure}

Prior to a large jump in the location of the slip event,  the lattice
has reached a state of strong pinning.  Hence,  following the analysis
presented in Ref. \cite{SVRIJMPC02}, if we condition the statistical
distribution of $\sigma_c(t)$ by the distance to the location of the
next slip event, along the $x$ direction for instance, $\Delta x$,  we
observe that the larger $\Delta x$, the narrower the distribution and
the closer its mean to the yield stress $\sigma^*$. These distributions
are shown in  Fig. \ref{stress-distribution}.  Motivated by 
the underlying criticality of the depinning transition,
we may anticipate a scaling
form of the distribution as
\begin{equation}
p(\sigma_c|\Delta x)=
\Delta x^{\nu_{\parallel}}
\psi \left[(\sigma^*-\sigma_c)\Delta x^{\nu_{\parallel}} \right]\;.
\label{stress-scaling}
\end{equation}

This particular form implies  that the standard deviation 
of the distribution, $\delta \sigma_c$  vanishes as $(\Delta
x)^{-\nu_\parallel}$, and that the mean value of
$\langle\sigma^*-\sigma_c\rangle(\Delta x)$ is simply proportional to 
$\delta\sigma_c(\Delta x)$.  The first property allows to determine
$\nu_{\parallel}$ and the second gives a simple way to estimate precisely $\sigma^*$
through a simple linear regression.  The same procedure applied to
$\Delta x$ gives a similar result.
Using the linear dependence of $\delta\sigma_c$ on $\sigma_c$
conditioned to the size of the activity jump in both the $x$ and $y$
direction, we find numerically $\sigma^*=0.517$ for a uniform distribution of threshold $\gamma$ in $[0,1]$ and a random slip amplitude from the same distribution.

The scaling of the standard deviation of the distribution versus the
jump size gives a determination of the exponents $\nu_\parallel\approx 0.68$ and
$\nu_\perp\approx 0.98$.  We note that again the ratio of these exponents 
gives the anisotropy scaling $\beta=\nu_\parallel/\nu_\perp=0.69$ in good
agreement with the previous determinations ($\beta\approx 0.65$).

The knowledge of the distribution $p(d)\propto d^{-\alpha_{\parallel}}$ of
the $x$-distances between successive active sites allows to express
the depinning stress distribution close to threshold:
\begin{eqnarray}
\label{universal}
\nonumber
{\cal Q}(\sigma^*-\sigma_c) &=&\int x^{\nu_\parallel-\alpha_\parallel} 
\psi \left[ (\sigma^*-\sigma_c)x^{\nu_\parallel} \right] dx
\\ &\propto &(\sigma^*-\sigma_c)^{\mu}  \;
\end{eqnarray}
where 
\be
 \mu=
{\frac{\alpha_{\parallel}-\nu_{\parallel}-1}{\nu_{\parallel}}}
\ee
The same argument obviously also holds for the $y$ direction.  This
latter scaling is also consistent with the anisotropy scaling $\beta=
(\alpha_\parallel-1)/(\alpha_\perp-1)\approx 0.64$.

 Despite its extreme simplicity, the model that we presented accounts
for several features of plasticity in amorphous materials. We could
identify a macroscopic yield stress. Below this threshold, the
material deforms plastically before blocking in jammed state. Above,
it can flow indefinitely. This behavior is typical of a
pinning/depinning situation. In the same spirit as the study presented
in Ref.\cite{SVRIJMPC02}, the model exhibits a critical behavior of
the plastic stress close to the macroscopic yield stress. When
submitted for the first time to a shear stress we observe a hardening
effect. In contrast with crystalline materials, here this effect
is of a pure statistical nature and corresponds  to a progressive
reinforcement of the weakest regions.  In addition to this global
hardening plastic behavior, the model exhibits a statistical localization. The
latter appears {\it via} elongated structures in the shear
direction. However, instead of concentrating onto a unique structure (such as
in Ref. \cite{SornettePRL93}), the plastic strain develops a complex
spatio-temporal organization. A statistical analysis of these patterns
reveals scaling properties; scaling exponents are summarized in table
\ref{tableexponents}.

Beyond this simplified model, the introduction of thermal
activation in the selection of the site to plastify should allow to
account for visco-plastic effects. Another improvement of such models
would consist in including both deviatoric and volumetric strain, the
latter coupling being characteristic of irreversible deformation in amorphous
solids.

\vfill

\begin{table}
\begin{center}
\begin{tabular}{|ccc|ccc|}\hline
$\zeta_\parallel$ &= &$-0.09 \pm 0.05$ &$\zeta_\perp$ &= &$0.50 \pm 0.05 $\\
\hline
$\alpha_\parallel$  &= &$1.61 \pm 0.05$ &$\alpha_\perp$  &= &$1.96 \pm 0.05$ \\
\hline
$\nu_\parallel$  &= &$0.68 \pm 0.05$ &$\nu_\perp$  &= &$0.98 \pm 0.05$ \\
\hline
$\mu$  &= &$0.00 \pm 0.05$ &$\beta$  &= &$0.65 \pm 0.05$\\ 
\hline
\end{tabular}
\end{center}
\caption{Table of scaling exponents}
\label{tableexponents}
\end{table}

\newpage

\vfill

\vfill

\newpage

\end{multicols}

\end{document}